\documentclass[12pt,a4paper]{amsart}
\usepackage[left=25mm,top=25mm,bottom=25mm,right=25mm]{geometry}
\usepackage[T1]{fontenc}
\usepackage{amssymb}
\usepackage[colorlinks=true,urlcolor=blue,citecolor=blue,linkcolor=blue,pdfstartview=FitH,pdfpagemode=None]{hyperref}

\newtheorem{theorem}{Theorem}[section]

\newtheorem{lemma}[theorem]{Lemma}
\newtheorem{proposition}[theorem]{Proposition}

\theoremstyle{definition}
\newtheorem{definition}[theorem]{Definition}

\theoremstyle{remark}
\newtheorem{remark}[theorem]{Remark}
\newtheorem{example}[theorem]{Example}

\newcommand{\A}{\mathcal{A}}
\newcommand{\NN}{\mathcal{N}}
\newcommand{\Real}{\mathbb{R}}
\newcommand{\Circle}{S^{1}}
\newcommand{\Cinf}{C^{\infty}}

\newcommand{\g}{\mathfrak{g}}
\newcommand{\gstar}{\mathfrak{g}^{*}}
\newcommand{\so}{\mathfrak{so}}
\newcommand{\Vect}{\mathrm{Vect}}
\newcommand{\Vectstar}{\mathrm{Vect}^{*}}
\newcommand{\SVect}{\mathrm{SVect}}
\newcommand{\SVectstar}{\mathrm{SVect}^{*}}

\newcommand{\SDiff}{\mathrm{SDiff}}
\newcommand{\SO}{\mathrm{SO}}

\newcommand{\norm}[1]{\left\Vert#1\right\Vert}
\newcommand{\abs}[1]{\left\vert#1\right\vert}
\newcommand{\set}[1]{\left\{#1\right\}}

\newcommand{\poisson}[2]{\{ #1,#2\,\}}
\newcommand{\dd}[2]{\frac{d{#1}}{d{#2}}}
\newcommand{\pd}[2]{\frac{\partial{#1}}{\partial{#2}}}
\newcommand{\pdd}[2]{\frac{\partial^{2}{#1}}{\partial{#2}^{2}}}
\newcommand{\pdl}[2]{\frac{\delta{#1}}{\delta{#2}}}
\newcommand{\pdlint}[2]{\frac{\delta^{\wedge}{#1}}{\delta{#2}}}
\newcommand{\pdlext}[2]{\frac{\delta^{\vee}{#1}}{\delta{#2}}}

\DeclareMathOperator{\grad}{\mathrm{grad}} %
\DeclareMathOperator{\curl}{\mathrm{curl}} %
\DeclareMathOperator{\dive}{div} %
\hypersetup{
    pdftitle={Poisson brackets in Hydrodynamics}, % Title
    pdfauthor={Boris Kolev}, % Author
    pdfsubject={2000 Mathematics Subject Classification: 53D20 53D17 37K05 37K65}, % Subject
    pdfkeywords={Hamiltonian structures, Hamiltonian systems on groups of diffeomorphisms}, % Keywords
    }

\begin{document}

\title[Poisson brackets in Hydrodynamics]{Poisson brackets in Hydrodynamics}%

% Author 1
\author[B. Kolev]{Boris Kolev}%
\address{CMI, 39 rue F. Joliot-Curie, 13453 Marseille Cedex 13, France}%
\email{boris.kolev@up.univ-mrs.fr}%

\thanks{I am very grateful to Jim Stasheff for the useful comments he sent me after the publication of this paper.}%


%\date{}%
%\dedicatory{}%
%\commby{}%

% ----------------------------------------------------------------

\begin{abstract}
This paper investigates different Poisson structures that have been proposed to give a Hamiltonian formulation to evolution equations issued from fluid mechanics. Our aim is to explore the main brackets which have been proposed and to discuss the difficulties which arise when one tries to give a rigorous meaning to these brackets. Our main interest is in the definition of a \emph{valid and usable} bracket to study \emph{rotational fluid flows with a free boundary}. We discuss some results which have emerged in the literature to solve some of the difficulties that arise. It appears to the author that the main problems are still open.
\end{abstract}

\maketitle

% ----------------------------------------------------------------

\section{Introduction}

The aim of this paper is to present a discussion of numerous attempts to use the Hamiltonian formalism of classical mechanics in hydromechanics and especially in the study of water waves. My motivation for this critical review came after a common work with David Sattinger \cite{KS06} and some discussions with Adrian Constantin about some of his work on water waves with vorticity \cite{CS04,CSW06}.

The interest for this subject goes back to Zakharov \cite{Zak68} who showed that \emph{irrotational} gravity waves could be given a \emph{Hamiltonian canonical structure}. It was also influenced by the success of the Hamiltonian formulation for one dimensional evolution equations such as the Korteweg-de Vries equation, a theme which has been extremely intensive in the seventies.

The Hamiltonian structure we refer to in this paper is that of general \emph{Poisson brackets} which gives a more general framework in the sense that Hamiltonian systems can be defined which are not necessarily \emph{canonical}. If this structure is well understood on finite dimensional manifolds, it is not the case for functional spaces. These structures have been defined at a \emph{formal} level, in the context of variational calculus \cite{GD79,GD81}. In infinite dimension, the brackets are not defined for all ``smooth functionals'' as it is the case for Poisson brackets on finite dimensional manifolds, but only for a subclass of such functionals. This leads therefore to two natural questions: is the bracket closed for the class of functionals on which it is defined and is the \emph{Jacobi identity}\footnote{The Jacobi identity is the fundamental equation which must be fulfilled by a Poisson bracket:
\begin{equation*}
    \poisson{\poisson{f}{g}}{h} + \poisson{\poisson{g}{h}}{f} + \poisson{\poisson{h}{f}}{g} = 0.
\end{equation*}} satisfied by this bracket ?

It appears that until recently, these questions have not been considered carefully and that Poisson brackets in functional spaces were defined \emph{up to boundary terms} as has been pointed out by Soloviev \cite{Sol97a} for instance.

It is however a fundamental question to check that a proposed bracket is a \emph{valid Hamiltonian structure} if one intends to go further than just a formal rewriting of the equations. To illustrate this fact I will just quote the work of Arnold \cite{Arn65} who was able to formulate a stability theorem for plane flows using a method now known as the \emph{Energy-Casimir method}. This work relies on the existence of \emph{Casimir functions}\footnote{A Casimir function is a smooth function whose bracket with every over smooth function vanishes.} for the underlying structure.

Besides, one could suggest that the ultimate goal of introducing Hamiltonian formalism in hydrodynamical problems and especially in the study of water waves would be to derive from it new results (like e.g. Arnold's stability theorem \cite{Arn65} or some recent results on particle trajectories
\cite{Con06,CE07}). To achieve this, one cannot however avoid the difficult question of defining a \emph{valid Hamiltonian structure}.

This paper proposes to discuss this question with a critical review of the main Hamiltonian structures which have been proposed in the literature (up to the author's knowledge).

The content of the paper is as follows. In Section~\ref{sec:Poisson_brackets}, we review the basic material on Poisson structures for finite dimensional manifolds. In Section~\ref{sec:Hamiltonian_structures}, we extend these definitions to functional spaces and raise the main difficulties which appear when one tries to define valid brackets in this more general context. In Section~\ref{sec:Vect_circle}, we discuss Hamiltonian structures on the space of smooth functions on the circle, where things work well. Section~\ref{sec:Arnold_bracket} is devoted to Arnold's bracket, a formulation of the Lie-Poisson bracket for the Lie algebra of divergence free vector fields on a compact domain and which is the background structure for the motions of an ideal fluid  with a fixed boundary. Several versions of this bracket are proposed and discussed. In the final section, Section~\ref{sec:free_boundary}, we discuss some brackets which were introduced in \cite{LMMR85} to study the difficult problem of fluids with vorticity and free boundary. It appears that the proposed bracket is not closed.

% ----------------------------------------------------------------

\section{Poisson brackets in finite dimension}\label{sec:Poisson_brackets}

\subsection{Symplectic and Poisson manifolds}

A \emph{symplectic manifold} is a pair $(M,\omega)$, where $M$ is a smooth manifold and $\omega$ is a \emph{closed} and \emph{nondegenerate} $2$-form on $M$. Such structures appear naturally in mechanics (see \cite{Kol04}). If $N$ is the \emph{configuration manifold} of a mechanical system, its phase space is the cotangent bundle $T^{*}N$ and is equipped with the canonical 2-form given by:
\begin{equation*}
    \sum_{i} dp_{i}\wedge dq^{i}.
\end{equation*}

Since a symplectic form $\omega$ is nondegenerate, it induces an
isomorphism $TM \to T^{*}M$. The inverse of this isomorphism defines a skew-symmetric bilinear form $P$ on the cotangent space $T^{*}M$ and a skew-symmetric bilinear mapping on $\Cinf(M)$, the space of smooth functions $f: M \to \mathbb{R}$, given by
\begin{equation}\label{equ:symplectic_bracket}
    \poisson{f}{g} = P(df,dg), \qquad f,g \in \Cinf(M) ,
\end{equation}
called the \emph{Poisson bracket} of the functions $f$ and $g$. For example, when $M = T^{*}N$ is a cotangent bundle, the corresponding bracket, known as the \emph{canonical bracket} is given by:
\begin{equation*}
    \poisson{f}{g} = \sum_{i} \pd{f}{p_{i}}\pd{g}{q^{i}} - \pd{g}{p_{i}}\pd{f}{q^{i}}
\end{equation*}

The observation that a bracket like \eqref{equ:symplectic_bracket} could be introduced on $C^\infty(M)$ for a smooth manifold $M$, without the use of a symplectic form, leads to the general notion of
a \emph{Poisson structure}.

\medskip

\begin{definition}
A \emph{Poisson structure} on a smooth manifold $M$ is a skew-symmetric
bilinear mapping $(f,g)\mapsto \poisson{f}{g}$ on the space $\Cinf(M)$,
which satisfies the \emph{Jacobi identity}
\begin{equation}\label{equ:Jacobi_identity}
    \poisson{\poisson{f}{g}}{h} + \poisson{\poisson{g}{h}}{f} +
    \poisson{\poisson{h}{f}}{g} = 0,
\end{equation}
as well as the \emph{Leibnitz identity}
\begin{equation}\label{equ:Leibnitz_identity}
    \poisson{f}{gh} = \poisson{f}{g}h + g\poisson{f}{h}.
\end{equation}
\end{definition}

Each Poisson bracket $\poisson{}{}$ corresponds to a smooth field $P$ of
\emph{bivectors}, called the \emph{Poisson bivector} of $(M,\poisson{}{})$ and such that
\begin{equation*}
    \poisson{f}{g}=P(df,dg),
\end{equation*}
for all $f,g \in \Cinf(M)$. The Jacobi identity implies that the bivector field $P$ must satisfy a certain condition, namely that $[P,P]=0$, where $[\; , \,]$ is the \emph{Schouten-Nijenhuis bracket}\footnote{The Schouten-Nijenhuis bracket is an extension of the Lie bracket of vector fields to skew-symmetric multivector fields, see~\cite{Vai94}.}.

The \emph{Hamiltonian vector field} of a smooth function $f$ on $M$ is defined by
\begin{equation*}
    X_{f} = P \,df
\end{equation*}
so that $\poisson{f}{h} = X_{h}\cdot f$. The Jacobi condition on $P$ insures that
\begin{equation*}
    X_{\poisson{f}{g}} = - [X_{f},X_{g}]
\end{equation*}
as in the symplectic case.

A \emph{Casimir function} is a smooth function $C$ on $M$ such that
\begin{equation*}
     \poisson{C}{f} = 0,\qquad \forall f \in \Cinf(M).
\end{equation*}
These functions play an important role in the study of the stability of equilibrium of Hamiltonian vector fields. Notice that in the symplectic case, the only Casimir functions are the constants.

\subsection{Poisson reduction}

Let us now explain how these Poisson structures appear naturally in mechanics. Let $N$ the \emph{configuration manifold} of a mechanical system and $M = T^{*}N$ its corresponding \emph{phase space}. It often happens that the system has some \emph{symmetries} represented by the (left) action of a Lie group $G$ on $N$. This action lifts to a \emph{symplectic action} of $G$ on $M = T^{*}N$, that is each diffeomorphism induced by an element $g\in G$ is a \emph{canonical transformation} of $M = T^{*}N$. If the group $G$ acts freely and properly on $M$, the \emph{reduced phase space} $M/G$ is a manifold and we may ask which structure from $M$ is inherited by the quotient space $M/G$.

For that purpose, let $\pi : M \to M/G$ be the canonical projection. Notice that $\ker \pi^{\prime}(x)$ is the tangent space to the $G$-orbit through $x$. Let $\omega$ be a 2-form on $M$, $P$ a bivector field on $M$ and recall the following criterions

\begin{enumerate}
  \item There exists a bivector field $\bar{P}$ on $M/G$ such that $\pi^{\prime} \circ P = \bar{P}\circ \pi$ if and only if
      \begin{equation*}
        (g^{*}P)(x) - P(x) \in \ker \pi^{\prime}(x)
      \end{equation*} for each point $x\in M$.
  \item There exists a 2-form $\bar{\omega}$ on $M/G$ such that $\pi_{*} \bar{\omega} = \omega$ if and only if
      \begin{equation*}
        g^{*}\omega = \omega \qquad \text{and} \qquad i_{X}\omega = 0
      \end{equation*}
      for each vector $X \in \ker \pi^{\prime}$.
\end{enumerate}

Notice that, unless $G$ is a discrete group, the second condition on $\omega$ is never satisfied and hence the symplectic structure on $M = T^{*}N$ cannot get down to $M/G$. However, condition (1) is fulfilled by the Poisson bivector $P$ of any Poisson structure on $M$ invariant under $G$, and leads naturally to the existence of a \emph{reduced Poisson structure} on $M/G$ such that $\pi : M \to M/G$ is a \emph{Poisson map}, i.e. such that
\begin{equation*}
    \poisson{f \circ \pi}{g \circ \pi} = \poisson{f}{g}\circ \pi
\end{equation*}
for all $f,g \in \Cinf(M/G)$. This process is known as the \emph{Poisson reduction} \cite{MW74}.

\subsubsection{Lie-Poisson structure}

The main illustration of this reduction process leads to the \emph{Lie-Poisson bracket}. Let $G$ be a Lie group and $\g$ its Lie algebra. The \emph{left action} on $G$ lift to a symplectic action on $T^{*}G\simeq G \times \gstar$ (equipped with the canonical symplectic structure) and induces a Poisson structure on $T^{*}G/G \simeq \gstar$ given by
\begin{equation}\label{equ:Lie_Poisson}
    \poisson{f}{g}(m) = - m([d_{m}f,d_{m}g])
\end{equation}
for $m\in\g^{*}$ and $f,g \in \Cinf(\g^{*})$\footnote{Here, $d_{m}f$, the differential of a function $f \in C^\infty(\g^{*})$ at $m \in \gstar$ is to be understood as an element of the Lie algebra $\g$.}. The corresponding Poisson bivector $P$ is given by
\begin{equation*}
    P_{m}(df,dg) = ad^{*}_{df}m(dg)
\end{equation*}
where $ad^{*}$ is the coadjoint action of $\g$ on $\gstar$.

\subsubsection{Euler equation}

The Lie-Poisson structure is the framework for the evolution equation known as the \emph{Euler equation} on a Lie group $G$. Consider a one-sided (left or right) invariant Riemannian metric $<\cdot,\cdot>$ on $G$. The geodesic flow corresponds to the flow of the Hamiltonian vector field on $T^{*}G$ equipped with the canonical structure and Hamiltonian
\begin{equation*}
    H(X_{g}) = \frac{1}{2}\, <X_{g},X_{g}>_{g}, \qquad X_{g}\in T^{*}G .
\end{equation*}
The reduced Hamiltonian function $H_{A}$ and the reduced Hamiltonian vector field $X_{A}$ on $\gstar$ are
\begin{equation*}
    H_{A}(m) = \frac{1}{2}\, (m,A^{-1}m), \qquad X_{A}(m) = ad^{*}_{A^{-1}m}m, \qquad m \in \gstar
\end{equation*}
where $A:u\mapsto <u,\cdot>_{e}$ is called the \emph{inertia operator}.

\begin{example}[The rigid body]
Euler equations of motion of a rigid body:
\begin{equation*}
    \dot{\omega}_{1} = \frac{I_{2}-I_{3}}{I_{1}} \omega_{2}\omega_{3}, \qquad \dot{\omega}_{2} = \frac{I_{3}-I_{1}}{I_{2}} \omega_{1}\omega_{3}, \qquad \dot{\omega}_{3} = \frac{I_{1}-I_{2}}{I_{3}} \omega_{1}\omega_{2}
\end{equation*}
are the basic example of Euler equations. In that case, the group $G$ is the rotation group $\SO(3)$. The Lie-Poisson bracket on $\so(3)^{*}\simeq \Real^{3}$ is given by
\begin{equation*}
    \set{f,g}(m) = m\cdot(\grad f(m)\wedge \grad g(m)), \qquad f,g \in \Cinf(\Real^{3}),
\end{equation*}
and the Hamiltonian is
\begin{equation*}
    H(m) = I_{1}^{-1}m_{1}^{2} + I_{2}^{-1}m_{2}^{2} + I_{3}^{-1}m_{3}^{2},
\end{equation*}
where $I_{1}, I_{2}, I_{3}$ are the \emph{principal moments of inertia} of the rigid body and $m_{k} = I_{k}\omega_{k}$.
\end{example}

% ----------------------------------------------------------------

\section{Poisson brackets in functional spaces}\label{sec:Hamiltonian_structures}

Several authors have tried to extend the notion of Poisson brackets to functional spaces in order to study evolution equations, see \cite{Olv93} for an excellent overview of the subject. There are however serious difficulties to handle when one enters into the details of these constructions as was pointed out in \cite{Sol93,Sol97b,Sol97a,Sol02a,Sol02b}.

In this section, we will review some of these difficulties. We consider Poisson brackets for smooth functionals defined on the the space $\Cinf(M)$ of smooth functions on a manifold $M$ or more generally on the space of smooth sections $\Gamma(E)$ of a vector bundle over $M$ (for simplicity, we will suppose that $M$ is the closure of an open subset of the Euclidean space $\Real^{n}$ with smooth boundary).

\subsection{Directional derivative \textit{versus} variational derivative}

Let $F$ be a smooth real function on some Fr\'{e}chet vector space $\Cinf(M,E)$ where $E$ is a finite dimensional vector space. The \emph{directional derivative} or \emph{Fr\'{e}chet derivative} of $F$ at $u$ in the direction $X\in \Cinf(M,E)$ is defined as
\begin{equation*}
    D_{X}F(u) = \left.\dd{}{\varepsilon}\right|_{\varepsilon=0} F(u + \varepsilon X).
\end{equation*}
In general, the directional derivative $X\mapsto D_{X}F(u)$ of a smooth functional $F$ is nothing more than a continuous linear functional on $\Cinf(M,E)$. Sometimes, this linear functional can be represented as
\begin{equation*}
    D_{X}F(u) = \int_{M} \pdl{F}{u}(u)\cdot X \, dV , \qquad \forall X \in \Cinf(M,E)
\end{equation*}
where
\begin{equation*}
    u \mapsto \pdl{F}{u}(u),
\end{equation*}
is a smooth map (vector field) from $\Cinf(M,E)$ to $\Cinf(M,E)$. The vector field $\delta F/\delta u$ is unique and we call it the $L^{2}$ \emph{gradient} of $F$.

\medskip

There is another notion of derivative, whose origin comes from variational calculus
\begin{equation*}
    DF(u). \delta u = \left.\dd{}{\varepsilon}\right|_{\varepsilon=0} F(u + \varepsilon \delta u)
\end{equation*}
where the \emph{variation} $\delta u$ has compact support and is \emph{subject to various boundary conditions}. We call it the \emph{variational derivative} of $F$. At first, it seems that the two definitions are the same. Of course, this is the case if $M$ is a compact manifold without boundary, but in general it is not.

A function $F$ on $\Cinf(M,E)$ is called a \emph{local functional} if
\begin{equation*}
    F(u) = \int_{M} f(x,u^{(r)})\, dV
\end{equation*}
depends of $u$ through a smooth function $f$ (the \emph{Lagrangian density} of $F$) which depends only on $x$ and the \emph{r-jet} of $u$ up to a certain order $r$. In that case, the Fr\'{e}chet derivative of $F$ is
\begin{equation*}
    D_{\delta u}F(u) = \left.\dd{}{\varepsilon}\right|_{\varepsilon=0} F(u + \varepsilon \delta u) = \int_{M} \sum_{J,k}\frac{\partial^{J}f}{\partial u_{k}^{J}}(x,u^{(r)})\,\delta u_{k}^{(J)}(x) \, dV
\end{equation*}
where $u_{1}, \dotsc ,u_{p}$ are the components of $u$ and
\begin{equation*}
    u_{k}^{(J)} = \frac{\partial^{\abs{J}}u_{k}}{\partial^{j_{1}}x^{1}\dotsb \partial^{j_{n}}x^{n}}, \qquad \abs{J} = j_{1}+\dotsb + j_{n}.
\end{equation*}
Using the Leibnitz rule repeatedly \cite{Olv93}, we can show that
\begin{equation*}
    \sum_{J,k}\frac{\partial^{J}f}{\partial u_{k}^{J}}\left(x,u_{k}^{(J)}(x)\right)\,\delta u_{k}^{(J)} = \sum_{k}\mathbf{E}_{k}(f)\,\delta u_{k} + \dive P
\end{equation*}
where
$\mathbf{E}_{k}$ is the \emph{Euler operator} defined by
\begin{equation*}
    \mathbf{E}_{k} = \sum_{J} (-D)_{J}\pd{}{u_{k}^{J}}, \qquad (-D)_{J} = (-D_{j_{1}})\dotsb(-D_{j_{n}}),
\end{equation*}
$P$ is a (functional) vector field
\begin{equation*}
    P(x,u^{(s)}) = \left(P_{1}(x,u^{(s)}), \dotsc , P_{n}(x,u^{(s)})\right)
\end{equation*}
and the divergence of $P$ is defined by
\begin{equation*}
    \dive P = D_{1}P_{1} + D_{2}P_{2} + \dotsb + D_{n}P_{n},
\end{equation*}
where $D_{i} = d/dx^{i}$ is the total derivative with respect to $x^{i}$.

Therefore, the  \emph{variational derivative} of a local functional $F$ can always be put in a gradient form
\begin{equation*}
    DF(u). \delta u = \int_{M} \delta F\cdot \delta u \, dV
\end{equation*}
where
\begin{equation*}
    \delta F = (\mathbf{E}_{1}(f), \dotsc , \mathbf{E}_{p}(f)).
\end{equation*}
However, when the manifold $M$ has non-empty boundary, the \emph{variational derivative} and the \emph{Fr\'{e}chet derivative} may differ by a boundary term. A local functional does not have necessarily a $L^{2}$ gradient relatively to its Fr\'{e}chet derivative.

\begin{example}
This may happen for instance for a local functional given by
\begin{equation*}
    F(u) = \int_{M} \dive P \, dV= \int_{\partial M} P \cdot n \,dS
\end{equation*}
The variational derivative of $F$ is identically zero but the Fr\'{e}chet derivative of $F$ has no reason to vanish and cannot be put into $L^{2}$ gradient form. This problem arises because in the definition of the Fr\'{e}chet derivative we allow all smooth variations whereas in the definition of the variational derivative we allow only variations subject to boundary conditions.
\end{example}

\medskip

A Poisson bracket $\poisson{F}{G}$ is first of all a bilinear map depending on the \emph{first derivative} of $F$ and $G$. Contrary to the finite dimensional case, it seems extremely difficult to define a \emph{tractable} Poisson bracket on the set of all functionals. The reasonable thing is to restrict the definition of the bracket to a subclass of functionals. For instance, in the \emph{formal variational calculus} \cite{Olv93}, a Poisson bracket is defined on the subclass of local functionals through a bilinear map on their variational derivatives but this bracket appears to be defined up to divergence terms. When the manifold is compact without boundary this may lead to a coherent Poisson bracket but when the manifold has non-empty boundary some difficulties arise.

\begin{example}[The Gardner bracket]
It was discovered by Gardner, \cite{Gar71}, that the Korteweg-de Vries equation
\begin{equation*}
    u_{t} = u_{xxx} + uu_{x}
\end{equation*}
can be written as a Hamiltonian equation using the bracket
\begin{equation*}
    \poisson{F}{G}(u) = \int_{\Circle} \pdl{F}{u} D_{x} \pdl{G}{u} \, dx .
\end{equation*}
and the Hamiltonian
\begin{equation*}
    H(u) =  \int_{\Circle} \left(-\frac{1}{2} u_{x}^{2} + \frac{1}{6}u^{3}\right)\, dx .
\end{equation*}
\end{example}

\subsection{Closure of the Poisson bracket and Jacobi identity}

As we have just seen, there is no well-defined Poisson bracket on the space of all smooth functionals. The known brackets are defined on a subclass $\A$ of functionals, called \emph{admissible} functionals.

When the manifold $M$ is compact without boundary, it is possible to choose for $\A$ the whole space of local functionals. We may then define a Poisson bracket $\poisson{F}{G}$ on $\A$ using an expression like
\begin{equation*}
    \poisson{F}{G} = \int_{M} \pdl{F}{u} P \pdl{G}{u} \, dV
\end{equation*}
where $P$ is a linear differential operator (witch may depend of the $r$-jet of $u$), as in the Gardner bracket. This gives us a well-defined bilinear map
\begin{equation*}
    \A \times \A \to \A
\end{equation*}
since the expression we have for $\poisson{F}{G}$ is itself a local functional.

When $M$ has non-empty boundary this is not sufficient and some other boundary conditions have to be introduced (see Section~\ref{sec:Arnold_bracket}). Now this leads to an immediate other question: If $F$ and $G$ satisfy this boundary conditions, is this true for $\poisson{F}{G}$? In other words is the class $\A$ of admissible functionals (verifying the boundary conditions) closed under the bracket ? As we shall see this is not at all obvious.

Finally and last but not least, if all these required conditions are satisfied, we have to check that the bracket verifies the Jacobi identity
\begin{equation*}
    \poisson{\poisson{F}{G}}{H} + \poisson{\poisson{G}{H}}{F} + \poisson{\poisson{H}{F}}{G} = 0 .
\end{equation*}
This last verification can be very tedious but the real difficulty remains however the closure of the bracket.

\subsection{Hamiltonian structures}

All these considerations lead us to introduce the following scheme to define a Poisson bracket on a functional space. First define a subspace $\A$ of smooth functionals (local functionals for instance, if $\partial M = \emptyset$, or local functionals with some boundary conditions otherwise). Then we introduce the following definition of a \emph{Hamiltonian structure}\footnote{The terminology \emph{Hamiltonian structure} is commonly used instead of \emph{Poisson structure} for functional spaces.} on $\A$.

\begin{definition}\label{defn:Hamiltonian_structure}
A \emph{Hamiltonian structure} on $\A$ is a bilinear operation $\poisson{\cdot}{\cdot}$ on $\A$ such that for any $F,G,H \in \A$ we have:
\begin{enumerate}
  \item $\poisson{F}{G} \in\A$,
  \item $\poisson{G}{F} = - \poisson{F}{G}$,
  \item $\poisson{\poisson{F}{G}}{H} + \poisson{\poisson{G}{H}}{F} + \poisson{\poisson{H}{F}}{G} = 0$.
\end{enumerate}
\end{definition}

\begin{remark}
Notice that the Leibnitz rule has been eliminated from the definition of a Hamiltonian structure. In fact, there is no well-defined commutative product on local functionals.
\end{remark}

In the following sections, we review some well-known brackets that have been proposed in the literature (see also \cite{Kol07,CK06}).

% ----------------------------------------------------------------

\section{The Lie-Poisson bracket on $\Vectstar(\Circle)$}\label{sec:Vect_circle}

In this section we will consider the Lie-Poisson bracket on the ``dual'' of the Lie algebra of smooth vector fields on the circle $\Vect(\Circle)\simeq\Cinf(\Circle)$. Recall that the canonical Lie-Poisson structure on the dual $\gstar$ of a Lie algebra $\g$ is given by
\begin{equation*}
    \poisson{F}{G}(m) = - m \left( [d_{m} F,d_{m} G] \right).
\end{equation*}
To give a sense to this expression, we have first to define an injection from $\g$ to $\gstar$.

\subsection{The regular dual}

Since the topological dual of the Fr\'{e}chet space $\Vect(\Circle)$ is too big
and not tractable for our purpose, being isomorphic to the space of
distributions on the circle, we restrict our attention in the
following to the \emph{regular dual} $\gstar$, the
subspace of $\Vectstar(\Circle)$ defined by linear functionals of the form
\begin{equation*}
    u \mapsto \int_{\Circle} mu \, dx,
\end{equation*}
for some function $m \in \Cinf(\Circle)$. The regular dual
$\gstar$ is therefore isomorphic to $\Cinf(\Circle)$
by means of the $L^2$ inner product\footnote{In the sequel, we use
the notation $u,v,\dotsc$ for elements of $\mathfrak{g}$ and
$m,n,\dots$ for elements of $\gstar$ to distinguish them,
although they all belong to $\Cinf(\Circle)$.}
\begin{equation*}
    <u , v > = \int_{\Circle} uv \, dx .
\end{equation*}

\subsection{Local functionals}

A local functional $F$ on $\Vectstar(\Circle)\simeq\Cinf(\Circle)$ is given by
\begin{equation*}
    F(m) = \int_{\Circle} f(x,m, m_{x}, \dotsc, m_{x}^{(r)}) \, dx .
\end{equation*}
Since there are no boundary terms, its \emph{functional derivative} $DF(m)$ is equal to its \emph{variational derivative}
\begin{equation*}
    DF(m).\delta m = \int_{\Circle} \pdl{F}{m}\, \delta m\, dx,\qquad m \in \Cinf(\Circle).
\end{equation*}
where
\begin{equation*}
    \pdl{F}{m} = \sum_{j=0}^{r} (-D_{x})^{j}\pd{f}{m^{j}}.
\end{equation*}
The map $m\mapsto \delta F/\delta m$ can be considered as a vector field on $\Cinf(\Circle)$, called the \emph{gradient} of $F$ for the $L^{2}$-metric. In other words, a local functional on $\Cinf(\Circle)$ has a smooth $L^{2}$ gradient.

\subsection{Hamiltonian structures on the regular dual}

To define a \emph{Poisson bracket} on the space of \emph{local functions} on $\Vectstar(\Circle)$, we consider a one-parameter family of linear operators $P_{m}$ ($m\in\Cinf(\Circle)$) whose coefficients are smooth function of $x$, $m$ and a finite number of its derivatives and set
\begin{equation}\label{equ:regular_Poisson_bracket}
    \poisson{F}{G}(m) = \int_{\Circle} \delta F \, P_{m} \, \delta G \, dx .
\end{equation}
where $\delta F$ and $\delta G$ stand here for the variational derivatives $\delta F/ \delta m$ and $\delta G/ \delta m$. The operators $P_{m}$ must satisfy certain conditions in order for
\eqref{equ:regular_Poisson_bracket} to be a valid Hamiltonian structure on the set $\A$ of local functionals on the regular dual $\Vectstar(\Circle)$. First it must be a skew-symmetric operator (relatively to the $L^{2}$ inner product).
\begin{equation*}
    \int_{\Circle} \delta F \, P_{m} \, \delta G \, dx  = - \int_{\Circle} \delta G \, P_{m} \, \delta F \, dx ,\qquad \forall F,G \in \A .
\end{equation*}
Since the expression for $\poisson{F}{G}$ is a local functional, the class of local functional is closed under this bilinear operation. Therefore we need only a criteria on $P$ to ensure that Jacobi identity is satisfied, in order to obtain a Hamiltonian structure.

\begin{lemma}\label{lem:Jacobi}
The Jacobi identity for \eqref{equ:regular_Poisson_bracket} is equivalent to the condition
\begin{equation}\label{equ:Jacobi_reformulation}
    \circlearrowleft \int_{\Circle} \delta F \, \left(D_{P\delta H}P\right)\delta G \, dx = 0
\end{equation}
for all $F,G,H \in \A$ where $\circlearrowleft$ indicates the sum
over cyclic permutations of $F,G,H$ and $ D_{\delta m}P$ is the Fr\'{e}chet derivative of $P$ in the direction $\delta m$.
\end{lemma}

\begin{remark}
Notice first that since $P$ is a linear differential operator whose coefficients are smooth functions of $x, m, m_{x}, \dotsc$, the Fr\'{e}chet derivative of $P$ in the direction $\delta m$ is just the linear differential operator obtained by replacing the coefficients of $P$ by their Fr\'{e}chet derivatives in the direction $X$. Since $P$ is assumed to be skew-symmetric, so is $D_{\delta m}P$.
\end{remark}

\begin{proof}
We already know that $\poisson{F}{G}$ is a local functional and hence its variational derivative $\delta \poisson{F}{G}$ is an $L^{2}$ gradient for $\poisson{F}{G}$, that is
\begin{equation*}
    D_{\delta m}\poisson{F}{G} = \int_{\Circle} \delta \poisson{F}{G} \, \delta m \, dx .
\end{equation*}
By definition of the bracket, we have
\begin{equation*}
    \poisson{\poisson{F}{G}}{H} =  \int_{\Circle} \delta \poisson{F}{G} P \delta H \, dx = D_{P \delta H}\poisson{F}{G}.
\end{equation*}
Using the fact that the second Fr\'{e}chet derivative is a symmetric operator and the fact that $P$ is skew-symmetric, we get
\begin{equation*}
    D_{\delta m} \poisson{F}{G} = \int_{\Circle} \left[\left(D_{\delta m}\delta F\right) P\delta G - \left(D_{P\delta F}\delta G\right) \delta m + \delta F \left(D_{\delta m}P\right)\delta G\right]\, dx
\end{equation*}
and hence
\begin{equation*}
    \poisson{\poisson{F}{G}}{H} = \int_{\Circle} \left[\left(D_{P\delta H}\delta F\right) P\delta G - \left(D_{P\delta F}\delta G\right) P\delta H + \delta F \left(D_{P\delta H}P\right)\delta G\right]\, dx .
\end{equation*}
Taking the sum over cyclic permutations of $F,G,H$, the two first terms of the right hand side of the last equation cancel and we obtain the equivalence of Jacobi identity with \eqref{equ:Jacobi_reformulation}, which ends the proof.
\end{proof}

To check \eqref{equ:Jacobi_reformulation} is still tedious in practice. Following Olver \cite{Olv93}, it is preferable to use the technique of \emph{functional bivectors}, which generalizes the notion of Poisson bivectors and Schouten-Nijenhuis brackets. First, given a functional density $f(x,m^{(r)})$, define
\begin{equation*}
    \theta (f) = f, \qquad \theta_{x} (f) = D_{x}f, \qquad \theta_{xx} (f) = D_{x}^{2}f, \qquad \dotsc
\end{equation*}
where $D_{x}$ stands for the total derivative relative to $x$. Extending the action of the differential operator $P$ on $\theta$ in an obvious way, we can write
\begin{equation*}
    \poisson{F}{G} = \frac{1}{2} \int_{\Circle} \left\{\theta(\delta F)(P\theta)(\delta G) - \theta(\delta G)(P\theta)(\delta F) \right\}\, dx = \frac{1}{2} \int_{\Circle} (\theta \wedge P\theta)(\delta F, \delta G) \, dx
\end{equation*}
so that
\begin{equation*}
    \Theta = \frac{1}{2} \int_{\Circle} \left\{\theta \wedge P\theta\right\}dx
\end{equation*}
appears as the analogue of the Poisson bivector for finite dimensional Poisson brackets.

\begin{example}
For the Gardner bracket we have
\begin{equation*}
\Theta = \frac{1}{2} \int_{\Circle} \left\{\theta \wedge \theta_{x}\right\}dx.
\end{equation*}
\end{example}

\begin{proposition}[Olver \cite{Olv93}]\label{prop:Olver}
A skew-symmetric linear differential operators $P$ (with coefficients depending on $x$, $m$, $m_{x}$, ...) defines a Hamiltonian structure on the space $\A$ of local functionals on $\Vectstar(\Circle)$ if and only if it satisfies
\begin{equation*}
    \int_{\Circle} \left\{\theta \wedge (D_{P\theta}P\wedge \theta)\right\}dx = 0.
\end{equation*}
\end{proposition}

\begin{remark}
Notice that the preceding expression is an alternatinng trilinear expression on functional densities. Note also that the two wedges have different meanings. The first one corresponds to wedging the ordinary multiplication of two functional densities whereas the second one is the wedging relative to the non-abelian bilinear operation $(f,g) \mapsto \left(D_{Pf}P\right)g$.
\end{remark}

\begin{proof}
Let $F,G,H$ be local functionals and $\delta F, \delta G$ and $\delta H$ their variational derivatives. Then
\begin{equation*}
    \frac{1}{2}\int_{\Circle} \left\{\theta \wedge (D_{P\theta}P\wedge \theta)\right\}(\delta F, \delta G, \delta H)dx = \, \circlearrowleft \int_{\Circle} \delta F \, \left(D_{P\delta H}P\right)\delta G \, dx .
\end{equation*}
Hence the proposition is just a corollary of lemma~\ref{lem:Jacobi}.
\end{proof}

\begin{example}
The Gardner bracket or more generally the bracket obtained from a skew-symmetric differential operator $P$ with constant coefficients satisfies the Jacobi identity since the Fr\'{e}chet derivative of such operators in any direction is zero and hence $D_{P\theta}P = 0$.
\end{example}

\begin{example}
The canonical Lie-Poisson structure on $\Vectstar(\Circle)$ is given by
\begin{equation}\label{equ:Lie_Poisson_circle}
    \poisson{F}{G}(m) = \int_{\Circle} m \left[ \delta F, \delta G\right] = - \int_{\Circle} \delta F \left( mD + Dm \right) \delta G\,dx
\end{equation}
It is represented by the skew-symmetric operator
\begin{equation*}
    P = -\left(mD + Dm\right) = - \left(2mD + m_{x}I\right)
\end{equation*}
where $D=d/dx$. We get
\begin{equation*}
    D_{P\theta}P = \left( 4m\theta_{x} + 2m_{x}\theta\right)D + \left(2m\theta_{xx} + 3m_{x}\theta_{x} + m_{xx}\theta\right)I .
\end{equation*}
hence
\begin{equation*}
    D_{P\theta}P\wedge \theta = 2m_{x}\theta \wedge \theta_{x} + 2m\theta_{xx} \wedge \theta + 3m_{x}\theta_{x} \wedge \theta
\end{equation*}
and
\begin{equation*}
    \theta \wedge (D_{P\theta}P\wedge \theta) = 0.
\end{equation*}
\end{example}

\begin{example}[Burgers equation]
The inviscid Burgers equation
\begin{equation*}
    u_{t} = - 3uu_{x}
\end{equation*}
can be written as an Euler equation on $\Vectstar(\Circle)$ with the Lie-Poisson bracket~\eqref{equ:Lie_Poisson_circle}. It corresponds to the inertia operator $m = Au = u$ and Hamiltonian
\begin{equation*}
    H(m) = \frac{1}{2}\int_{\Circle} m^{2} \, dx .
\end{equation*}
\end{example}

\begin{example}[Camassa-Holm equation]\label{exple:Camassa_Holm}
The Camassa-Holm equation \cite{CH93}
\begin{equation*}
    u_{t} - u_{txx} + 3uu_{x} - 2u_{x}u_{xx} - uu_{xxx} = 0
\end{equation*}
can be written as an Euler equation on $\Vectstar(\Circle)$ with the Lie-Poisson bracket~\eqref{equ:Lie_Poisson_circle}. It corresponds to the inertia operator $m = Au = u-u_{xx}$ and Hamiltonian
\begin{equation*}
    H(m) = \frac{1}{2}\int_{\Circle} mu \, dx ,
\end{equation*}
cf. \cite{Mis98} - see also the discussion in \cite{CK03}.

Notice however that $H$ is not a local functional of $m$ since it depends on $m$ by the intermediary of the non local operator $A^{-1}$. To overcome this difficulty, one may try to extend the Hamiltonian structure~\eqref{equ:Lie_Poisson_circle} for functionals which are local expressions $x, u, u_{x}, m, m_{x}, \dotsc $ where $u = A^{-1}m$ rather than $x, m, m_{x}, \dotsc $. But this space of functionals is not closed under the preceding bracket and the space of functionals generated by successive brackets of such functionals seems tedious to describe.

In that case however, it is possible to overcome these difficulties by extending the Hamiltonian structure to the whole space of \emph{smooth functionals} which have a $L^{2}$ \emph{smooth gradient}, that is
\begin{equation*}
    D_{\delta m}F(m) = \int_{\Circle}\delta F(m) \, \delta m \, dx
\end{equation*}
where $m \mapsto \delta F(m)$ is a smooth smooth map from $\Cinf(\Circle)$ to $\Cinf(\Circle)$. The first two axioms which define a Hamiltonian structure are verified. Indeed, the bracket of two such functionals has itself a smooth gradient, namely
\begin{equation*}
    \delta \poisson{F}{G} = D_{P \delta F}\, \delta G - D_{P \delta G}\,\delta F + \delta G \, D_{x}\delta F - \delta F \, D_{x}\delta G .
\end{equation*}
Finally, Jacobi identity is also verified for this extension. In fact, Lemma~\ref{lem:Jacobi} and Proposition~\ref{prop:Olver} are still valid for those more general functionals.
\end{example}

% ----------------------------------------------------------------

\section{Poisson brackets for ideal fluids in a fixed domain}
\label{sec:Arnold_bracket}

Let $\Omega$ be a relatively compact domain in $\Real^{2}$ or $\Real^{3}$ with a smooth boundary. We let $\SDiff(\Omega)$ be the group of volume-preserving smooth diffeomorphisms of $\Omega$ and $\SVect(\Omega)$ the Lie algebra of divergence-free vector fields on $\Omega$, tangent to the boundary, which can be interpreted as the Lie algebra of $\SDiff(\Omega)$. In a famous article \cite{Arn66}, Arnold showed that the Euler equations of perfect incompressible fluid flows in a fixed domain
\begin{equation}\label{equ:ideal_fluids}
    \pd{u}{t} + \nabla_{u}u = - \grad p, \qquad \dive u = 0, \qquad u\cdot n = 0 \text{ on } \partial \Omega
\end{equation}
could be interpreted as a the Euler equation of the right-invariant (weak) Riemannian metric
\begin{equation}\label{equ:L2_metric}
    <u_{\varphi},v_{\varphi}> = \int_{\Omega} u_{\varphi} \cdot v_{\varphi} \, dV
\end{equation}
where $u_{\varphi},v_{\varphi}$ are vector fields over $\varphi\in\SDiff(\Omega)$ (\emph{Lagrangian} velocities).

\medskip

The \emph{regular dual} of $\SVect(\Omega)$, noted $\SVectstar(\Omega)$, consists of linear functionals on $\SVect(\Omega)$ with smooth density $\alpha\in\Omega^{1}(\Omega)$
\begin{equation*}
    u \mapsto \int_{\Omega} \alpha(u)\, dV .
\end{equation*}
Since exact one-forms are $L^{2}$-orthogonal to divergence-free vector fields, a one-form $\alpha\in \Omega^{1}(\Omega)$ represents an element of $\SVectstar(\Omega)$ only up to total differential. Each element of $\SVectstar(\Omega)$ is therefore represented by a class $[\alpha]$ in $\Omega^{1}(\Omega)/d\Omega^{0}(\Omega)$. If moreover, $\Omega$ is a simply connected domain, $\Omega^{1}(\Omega)/d\Omega^{0}(\Omega)$ is isomorphic to $d\Omega^{1}(\Omega)$ via the exterior derivative $d$ and the class $[\alpha]$ is completely represented by the two-form $\omega = d\alpha$, called the \emph{vorticity}.

\subsection{Arnold bracket}

The bracket, now known as \emph{Arnold bracket}, is defined for smooth functionals $F$ on $\SVectstar(\Omega)$ whose Fr\'{e}chet derivative can be written as
\begin{equation}\label{equ:first_gradient}
    D_{[\delta \alpha]}F(\omega) = \int_{M} \delta \alpha\left(\pdl{F}{\omega}\right) \, dV \qquad \text{where} \qquad \pdl{F}{\omega}\in\SVect(\Omega),
\end{equation}
in other words, for those functionals which have a $L^{2}$ gradient in the Lie algebra $\SVect(\Omega)$. It is given by the formula\footnote{The equality of the two formulations results from
\begin{equation*}
    d\alpha(u,v) = u \cdot \grad \alpha(v) - v \cdot \grad \alpha(u) - \alpha([u,v]).
\end{equation*}
}
\begin{equation}\label{equ:Arnold_bracket}
    \poisson{F}{G}(\omega) = - \int_{M}\alpha\left(\left[\pdl{F}{\omega},\pdl{G}{\omega}\right]\right)\, dV = \int_{M} \omega\left(\pdl{F}{\omega},\pdl{G}{\omega}\right)\, dV .
\end{equation}

\begin{proposition}
The bracket defined by equation~\eqref{equ:Arnold_bracket} is a valid Hamiltonian structure on the space of smooth functionals on $\SVectstar(\Omega)$ which have a smooth gradient in $\SVect(\Omega)$.
\end{proposition}

\begin{proof}
We have to check the three properties of Definition~\ref{defn:Hamiltonian_structure}. Expression~\eqref{equ:Arnold_bracket} is clearly skew-symmetric. To show that the bracket is closed, we recall first that the symmetry of the \emph{second Fr\'{e}chet derivative} leads to
\begin{equation*}
    \int_{\Omega} \delta \alpha \left( D_{[\delta \beta]} \delta F\right) \, dV = \int_{\Omega} \delta \beta \left( D_{[\delta \alpha]} \delta F\right) \, dV
\end{equation*}
for every admissible functional $F$. From this property, we deduce that for every admissible function $F$ and $G$, we have
\begin{equation*}
    D_{[\delta \alpha]}\poisson{F}{G}(\omega) = \int_{\Omega} \delta \alpha \left( D_{[i_{\delta F}\omega]}\delta G - D_{[i_{\delta G}\omega]}\delta F - \left[\delta F, \delta G\right] \right)\, dV,
\end{equation*}
that is $\poisson{F}{G}$ is also admissible with gradient
\begin{equation*}
    \delta \poisson{F}{G} = D_{[i_{\delta F}\omega]}\delta G - D_{[i_{\delta G}\omega]}\delta F - \left[\delta F, \delta G\right].
\end{equation*}
It remains to check Jacobi identity. We can write
\begin{equation*}
    \poisson{F}{G}(\omega) = \int_{\Omega} \delta F(\omega) \cdot P_{\omega}\,\delta G(\omega) \, dV
\end{equation*}
where
\begin{equation*}
    P_{\omega}:\SVect(\Omega) \to \SVect(\Omega), \qquad u \mapsto -\mathrm{Proj}\left(i_{u}\omega\right)
\end{equation*}
and $\mathrm{Proj}$ is the projection from $\Omega^{1}(\Omega)$ onto $\SVect(\Omega)$, which associates to a one-form $\alpha$ the unique divergence free vector field $v$, tangent to the boundary, such that
\begin{equation*}
    \int_{\Omega} \alpha(w)\, dV = \int_{\Omega} v\cdot w\, dV, \qquad \forall w\in\SVect(\Omega).
\end{equation*}
With these notations, we get as in Lemma~\ref{lem:Jacobi}
\begin{align*}
    \circlearrowleft \poisson{\poisson{F}{G}}{H}  & = - \circlearrowleft \int_{\Omega} [\delta F,\delta G]\cdot P\delta H  \, dV \\
    & = \,\circlearrowleft \int_{\Omega} \omega(\delta H,[\delta F,\delta G]) \, dV \\
    & = - \circlearrowleft \int_{\Omega} \alpha([\delta H,[\delta F,\delta G]]) \, dV = 0
\end{align*}
where $\omega = d\alpha$.
\end{proof}

\subsubsection{Euler-Helmholtz equation}

Arnold's bracket~\eqref{equ:Arnold_bracket} can be used to interpret Euler's equations of perfect incompressible fluid flows~\eqref{equ:ideal_fluids} in their \emph{Helmholtz} or \emph{vorticity} representation
\begin{equation}\label{equ:Helmholtz}
    \partial_{t}\omega = \curl(u\times \omega), \qquad \omega = \curl u
\end{equation}
as the Euler equation of the $L^{2}$ metric~\eqref{equ:L2_metric}.

Recall that the curl of a vector field $u$ is defined as the unique vector $\omega$ such that
\begin{equation*}
    i_{\omega}\,vol = du^{\flat}
\end{equation*}
where $u^{\flat}$ is the covariant representation of $u$. Therefore, $\SVectstar(\Omega)$, the space of exact two-forms can be identified with the space of curls and the inertia operator of the $L^{2}$ metric~\eqref{equ:L2_metric} can be described as
\begin{equation*}
    A : \SVect(\Omega) \to \SVectstar(\Omega), \qquad u \mapsto \curl u .
\end{equation*}
This operator is invertible. Let $\omega\in\SVectstar(\Omega)$ be a curl. Then $u=A^{-1}\omega$ is the unique solution of the problem
\begin{equation*}
    \curl u = \omega, \qquad \dive u = 0, \qquad u\cdot n = 0\text{ on } \partial \Omega .
\end{equation*}
The Hamiltonian is given by
\begin{equation*}
    H(\omega) = \frac{1}{2} \int_{\Omega} \norm{u^{2}}\, dV , \qquad u = A^{-1}\omega .
\end{equation*}
We have
\begin{equation*}
    D_{[\delta \alpha]}H(\omega) = \int_{\Omega} u\cdot \delta u \, dV, \qquad \delta \alpha = (\delta u)^{\flat}
\end{equation*}
and hence $H$ is an admissible functional with gradient
\begin{equation*}
    \delta H(\omega) = u = A^{-1}\omega .
\end{equation*}
Euler equation $\dot{F} = \poisson{F}{H}$, for all admissible functional\footnote{Each vector field $u\in\SVect(\Omega)$ can be realized as the gradient of an admissible functional, namely of the linear functional
\begin{equation*}
    F(\omega) = \int_{\Omega} \alpha(u)\, dV,\qquad d\alpha = \omega .
\end{equation*}
} $F$ gives
\begin{equation*}
    \int_{\Omega} \partial_{t}u\cdot \delta F \, dV = \int_{\Omega} \omega\cdot (\delta F \times u)  \, dV = \int_{\Omega} \delta F \cdot (u \times \omega)  \, dV,
\end{equation*}
that is
\begin{equation*}
    \partial_{t}u = u \times \omega,\quad \text{modulo a gradient}
\end{equation*}
and taking the curl, we get
\begin{equation*}
    \partial_{t}\omega = \curl(u \times \omega).
\end{equation*}

\begin{remark}
We could have restricted the definition of Arnold's Poisson bracket for local functionals which have a $L^{2}$ gradient. In fact this space is closed under the bracket. But this would not have permitted us to treat the hydrodynamic problem since the Hamiltonian is not a local functional (see Example~\ref{exple:Camassa_Holm} for a similar situation in dimension 1).
\end{remark}

\begin{remark}
In several papers, the Arnold bracket is written as
\begin{equation*}
    \poisson{F}{G}(u) =  \int_{\Omega} \curl u \cdot \left(\pdl{F}{u}\times \pdl{G}{u}\right)\, dV .
\end{equation*}
for smooth functionals with smooth $L^{2}$ gradient on the Lie algebra $\SVect(\Omega)$ rather than $\SVectstar(\Omega)$. This is just the ``pullback'' of \eqref{equ:Arnold_bracket} by the inertia operator $A$. The fact that this bracket preserves the space of functionals which have a $L^{2}$ gradient is less obvious to see in this expression because of the term $\curl u$ which leads to an integration by parts, but in fact it works. The advantage of using variables $u$ instead of $\omega$ is that the Hamiltonian becomes a local functional in these variables. In that case, the Hamiltonian equation, $\dot{F} = \poisson{F}{H}$ for all admissible $F$, leads directly to equations~\eqref{equ:ideal_fluids}.
\end{remark}

\begin{remark}\label{rem:Olver_formulation}
A third interpretation of Arnold bracket was given in \cite{Olv93}. It is defined, in the context of \emph{formal variational calculus} (where boundary terms are ignored) for local functionals on $\SVectstar(\Omega)$. The gradient of a functional $F$ is defined here as
\begin{equation*}
    DF(\omega).\delta \omega = \int_{\Omega} \pdl{F}{\omega} \cdot \delta \omega \, dV
\end{equation*}
where the gradient, $\delta F / \delta \omega$ is a divergence free vector field and the variation $\delta \omega$ is assumed to vanish on the boundary. Notice that the definition of the gradient given here is quiet different from the previous definition~\eqref{equ:first_gradient}. Indeed the two definitions differ through a boundary term
\begin{equation*}
    \int_{\Omega} \pdl{F}{\omega} \cdot \delta \omega \, dV = \int_{\Omega} \delta \alpha \left(\curl \pdl{F}{\omega}\right) \, dV + \int_{\partial \Omega} \left( \pdl{F}{\omega} \times \delta u \right)\cdot n \, dS
\end{equation*}
where $\delta \omega = \curl \delta u$ and $\delta \alpha = (\delta u)^{\sharp}$. Therefore we cannot conclude that both Poisson structure are rigorously equivalent.

For two-dimensional flows, the Hamiltonian operator $P$ is represented as
\begin{equation*}
    P = \omega_{x} D_{y} - \omega_{y} D_{x}
\end{equation*}
and the gradient of the Hamiltonian
\begin{equation*}
    H(\omega) = \frac{1}{2} \int_{\Omega} \norm{u^{2}}\, dV , \qquad u = A^{-1}\omega .
\end{equation*}
$\delta H / \delta \omega$ is the \emph{stream function} $\psi$ of the velocity $u$. It was shown in \cite{Olv93} that in this context, the Jacobi identity was satisfied and that the Hamiltonian equation $\dot{F} = \poisson{F}{H}$ was equivalent to Euler-Helmholtz equation~\eqref{equ:Helmholtz}. We insist on the fact that the computations which leads to these results relies on the \emph{vanishing of the variations} on the boundary.
\end{remark}

\subsubsection{Enstrophy}

This Poisson bracket \eqref{equ:Arnold_bracket} has been rejected by the authors in \cite{LMMR85} because for two-dimensional flows, the \emph{generalized enstrophy} functional
\begin{equation*}
    C(\omega) = \int_{\Omega} \phi(\omega) \, dx\wedge dy
\end{equation*}
which is known to be invariant under the coadjoint action of $\SDiff(\Omega)$ on $\SVectstar(\Omega)$ is not a \emph{Casimir function} for this bracket.
Indeed
\begin{equation*}
    D_{[\delta \alpha]}C(\omega) = \int_{\Omega} \delta \alpha\big(\curl (\phi^{\prime}(\omega)\hat{k})\big)\,dx\wedge dy +\oint_{\partial \Omega} \phi^{\prime}(\omega)\delta \alpha
\end{equation*}
has some boundary terms and is therefore \emph{not an admissible functional} for~\eqref{equ:Arnold_bracket}.

\subsection{Second LMMR bracket}
\label{subsec:second_LMMR}

Since Casimir functions play a fundamental role in the study of stability of two-dimensional flows as it has been shown in \cite{Arn65}, the authors in \cite{LMMR85} have proposed to improve the definition of Arnold's bracket by taking into account boundary terms so that the enstrophy becomes a Casimir function.

They have derived this bracket using the same reduction process which has been used for Arnold's bracket. The difference lies in a different choice of admissible functionals.

The starting point is the \emph{Lagrangian} description of the problem. For an incompressible fluid moving in a fixed domain $\Omega$, the \emph{configuration space} is the group of volume-preserving diffeomorphisms $\SDiff(\Omega)$. The \emph{phase space}, $T^{*}\SDiff(\Omega)$ has to be understood as the set of pairs $(\varphi,\mu)$ where $\varphi\in\SDiff(\Omega)$ is the ``base point'' and $\mu$ is a one-form over $\varphi$ (i.e. for each $x$, $\mu(x)\in T^{*}_{\varphi(x)}\Omega$).

The class of admissible functionals $F$, previously limited to smooth functionals which have smooth $L^{2}$ gradient, is now extended to ones whose ``gradients'' can be written as
\begin{equation*}
    \pdl{F}{\varphi} = \pdlint{F}{\varphi} + \delta_{\partial \Omega}\pdlext{F}{\varphi}, \qquad \pdl{F}{\mu} = \pdlint{F}{\mu} + \delta_{\partial \Omega}\pdlext{F}{\mu},
\end{equation*}
where $\delta_{\partial \Omega}$ is the Dirac measure on $\Omega$ concentrated on $\partial \Omega$~\footnote{Notice however that this decomposition is not unique.}. A Poisson bracket can be defined for those functionals using the \emph{formal canonical bracket} on $T^{*}\SDiff(\Omega)$
\begin{equation*}
    \poisson{F}{G} = \int_{\Omega} \left( \pdl{F}{\varphi}\pdl{G}{\mu}-\pdl{G}{\varphi}\pdl{F}{\mu}\right)\,dV
\end{equation*}
provided that the boundary condition
\begin{equation}\label{equ:square_deltas}
    \pdlext{F}{\varphi}\pdlext{G}{\mu}-\pdlext{G}{\varphi}\pdlext{F}{\mu} =0
\end{equation}
is satisfied to avoid squares of delta functions.

The Lie-Poisson reduction of the phase space $T^{*}\SDiff(\Omega)$ by the gauge group $\SDiff(\Omega)$ (corresponding to relabeling fluid particles) leads to the Second LMMR bracket \cite{LMMR85} defined for functionals on $\Vectstar(\Omega)$ whose Fr\'{e}chet derivative can be written as
\begin{equation*}
    DF(u).\delta u = \int_{\Omega} \pdlint{F}{u} \cdot \delta u \, dV +  \int_{\partial\Omega} \pdlext{F}{u} \cdot \delta u \, dS
\end{equation*}

The expression for the resulting bracket is quite complicated and will not be given here. It must be stated, however that this bracket is well-defined for a pair of admissible functionals $(F,G)$ only if condition~\eqref{equ:square_deltas} is satisfied. This will be the case if one if one of the functionals $F$ or $G$ satisfy $\delta^{\vee}F/\delta u = 0$. It was shown in \cite{LMMR85} that for two-dimensional flows, the generalized enstrophy was a Casimir function for this bracket in the sense that $\poisson{C}{F} = 0$ for all functions admissible function $F$ such that $\delta^{\vee}F/\delta u = 0$.

We will not try to check that this second LMMR bracket is a valid Hamiltonian structure. The definition of this bracket $\poisson{F}{G}$ requires the condition~\eqref{equ:square_deltas} on the pair of functionals $(F,G)$ to be satisfied. But this latest condition concerns the pair $(F,G)$ and not each of the functionals $F,G$ alone. Therefore, it is not clear on which subclass of functionals is this bracket defined.

\subsection{Soloviev bracket}

In a series of papers, \cite{Sol93,Sol02a,Sol02b}, Soloviev tried to define a Poisson bracket for local functionals which avoids this tedious boundary condition~\eqref{equ:square_deltas}. The idea introduced in \cite{Sol93} is to define a bracket involving not only the "first gradient" (the factor of $\delta u$) but the complete set of "higher order gradients" (the factor of $(\delta u)^{(J)}$) of a local functional.

Using the Leibnitz rule but making no integration by parts, we can write the Fr\'{e}chet derivative of a local functional $F$ as
\begin{equation*}
    DF(u).\delta u = \int_{\Omega} \sum_{J,k}\frac{\partial^{J}f}{\partial u_{k}^{J}}\left(x,u^{(r)}\right)\,\delta u_{k}^{(J)}(x) \, dV = \int_{\Omega} \sum_{J,k} D_{J}\left( E^{J}_{k}(f) \delta u_{k}\right)\, dV
\end{equation*}
where the higher Eulerian operators\footnote{Notice that all the sums are finite since only a finite number of derivatives appear in all these formula. The zero order higher Eulerian operator $E^{0}_{k}$ is just the ordinary Euler operator $E_{k}$.} $E^{J}_{k}$ are defined by
\begin{equation*}
    E^{J}_{k}(f) = \sum_{K\supset J}
    \left(
      \begin{array}{c}
        K \\
        J \\
      \end{array}
    \right)
    (-D)_{K\setminus J} \pd{f}{u_{k}^{(K)}},
\end{equation*}
the binomial coefficients for multi-indices are
\begin{equation*}
    \left(
      \begin{array}{c}
        K \\
        J \\
      \end{array}
    \right) =
    \left(
      \begin{array}{c}
        k_{1} \\
        j_{1} \\
      \end{array}
    \right) \dotsb
    \left(
      \begin{array}{c}
        k_{r} \\
        j_{r} \\
      \end{array}
    \right)
\end{equation*}
and
\begin{equation*}
    (-D)_{K} = (-1)^{\abs{K}}D_{K}.
\end{equation*}
The following formula was derived by Soloviev in \cite{Sol93} to define a Poisson bracket on the class of all local functionals
\begin{equation*}
    \poisson{F}{G} = \sum_{J,K}\sum_{p,q}\int_{\Omega}D_{J + K}\left(E^{J}_{p}(f)I_{pq}E^{K}_{q}(g)\right)\, dV
\end{equation*}
where the operator $I_{pq}$ are subject to certain conditions to satisfy Jacobi identity.

\begin{example}
In \cite{Sol02a}, this method was illustrated for the formulation of Arnold's bracket presented in Remark~\ref{rem:Olver_formulation} for 2 dimensional flows. The antisymmetric operator $I$ was given in this case by
\begin{equation*}
    I = \theta(\omega_{x}D_{y} - \omega_{y}D_{x}) + \frac{1}{2} (D_{y}\theta\omega_{x}-D_{x}\theta \omega_{y}),
\end{equation*}
where the derivative of the characteristic function $\theta = \theta_{\Omega}$ has to be understood in the sense of distributions using certain rules \cite{Sol02a}. It was shown that, up to these rules, we obtain a valid Poisson structure. There is however one objection on this example: up to my understanding, Soloviev's formalism was developed for local functionals but the Hamiltonian giving rise to the Euler equations in this case is
\begin{equation*}
    H(\omega) = \frac{1}{2}\int_{\Omega} \norm{u}^{2}\, dS, \qquad \omega =\curl u,
\end{equation*}
which is not a local functional of the variable $\omega$.
\end{example}

% ----------------------------------------------------------------

\section{Poisson brackets for ideal fluids with a free boundary}\label{sec:free_boundary}

In 1968, Zakharov \cite{Zak68} showed that Euler's equations for \emph{irrotational gravity waves} could be written as a \emph{canonical Hamiltonian system}. The Hamiltonian is
\begin{equation*}
    H = \frac{1}{2} \iiint \limits_{D}(\grad\varphi)^{2}\,dV + \frac{1}{2}\lambda\iint\limits_{\Real^2}\zeta^{2}(x,y,t)\,dS.
\end{equation*}
The Poisson brackets implicit in Zakharov's observation are the
canonical brackets
\begin{equation*}
    \poisson{F}{G} = \iint\limits_{\Real^2} \left(\pdl{F}{\varphi}\pdl{G}{\zeta} - \pdl{ F}{\zeta}\pdl{G}{\varphi}\right)dS;
\end{equation*}
the Hamiltonian flow is then the canonical flow
\begin{equation*}
    \zeta_{t} = \pdl{H}{\varphi}, \qquad \varphi_{t} = -\pdl{H}{\zeta}.
\end{equation*}

The Hamiltonian $H$ is regarded as a functional of
$(\widetilde{\varphi},\,\zeta)$ where $\zeta=\zeta(x,y,t)$ is the height of
the free surface, and $\widetilde{\varphi}=\varphi\vert_{\partial D}$ is the trace of
the harmonic function $\varphi$ on the free surface, with
$\partial_{n}\varphi = 0$ on the bottom. The evolution takes place in the
space of harmonic functions on $D$. Zakharov's result is verified by calculating the gradients of $H$
with respect to $\zeta$ and $\varphi$.

In \cite{LMMR85}, a generalization of this Hamiltonian structure was proposed for incompressible fluid flows with possible vorticity. It is however no longer a canonical structure. The approach used in \cite{LMMR85} to derive a Hamiltonian structure is essentially the same as the one used to derive Arnold's bracket (fixed domain): using a Poisson reduction process of the canonical symplectic structure on the phase space by the gauge group (relabelling of particles). The main difference is that in the \emph{free boundary} case, the gauge group \emph{no longer acts transitively} on the configuration space (the space of embeddings of a reference domain in $\Real^{n}$).

\subsection{First LMMR bracket}

This structure, known as the first LMMR bracket is defined on the space of pairs $(v,\Sigma)$, where $\Sigma$ is the free surface and $v$ is the spatial velocity field, a divergence free vector field defined on $D_{\Sigma}$, the region bounded by $\Sigma$. The surface $\Sigma$ is assumed to be compact and diffeomorphic to the boundary of a reference region $D$.

The class $\A$ of functionals $F : \NN \to \Real$ on which this bracket is defined is formed by functionals with the following properties:

\begin{enumerate}
  \item A variation $\delta v$ is just a divergence free vector field on $D_{\Sigma}$ and we assume that there exists a divergence free vector field $\delta F / \delta v$ defined on $D_{\Sigma}$ such that for all variations $\delta v$:
  \begin{equation*}
    D_{v}F(v,\Sigma)\cdot\delta v = \int_{D_{\Sigma}} \pdl{F}{v}\cdot \delta v\,dV
  \end{equation*}
  where $D_{v}F$ is the derivative of $F$ holding $\Sigma$ fixed.
  \item A variation $\delta \Sigma$, which is a function on $\Sigma$, has to be understand as an infinitesimal variation of $\Sigma$ in its normal direction. Since only volume preserving variations are allowed, $\delta \Sigma$ has zero integral over $\Sigma$. We assume that there exists a smooth function $\delta F / \delta \Sigma$ such that for all variations $\delta \Sigma$:
  \begin{equation*}
    D_{\Sigma}F(v,\Sigma)\cdot \delta \Sigma = \int_{\Sigma} \pdl{F}{\Sigma} \delta \Sigma\,dS
  \end{equation*}
  where $D_{\Sigma}F$ is the derivative of $F$ holding $v$ constant~\footnote{This definition requires us to extend smoothly $v$ in a neighborhood of $\Sigma$. One can check that $\delta \Sigma$ is independent on the way $v$ is extended and that it is determined up to an additive constant.}.
\end{enumerate}

The Poisson bracket on functions $F,G\in \A$ is defined by
\begin{equation*}
    \poisson{F}{G} = \int_{D_{\Sigma}} \omega \cdot \left( \pdl{F}{v} \times \pdl{G}{v}\right) \, dV + \int_{\Sigma} \left(\pdl{F}{\Sigma}\pdl{G}{\phi} - \pdl{G}{\Sigma}\pdl{F}{\phi} \right)\,dS
\end{equation*}
where $\omega = \curl v$ and
\begin{equation*}
    \pdl{F}{\phi} = \left.\pdl{F}{v}\right|_{\Sigma}\cdot n.
\end{equation*}
This last term corresponds to the variational derivative of $F$ taken with respect to variations of $v$ by potential flows.

It has been checked in \cite{LMMR85} that the Hamiltonian equation $\dot{F} = \poisson{F}{H}$ is equivalent to the equations of a liquid drop
\begin{equation*}
    \pd{v}{t} + \nabla_{v}v = - \grad p, \qquad \pd{\Sigma}{t} = v \cdot n, \qquad \dive v = 0, \qquad p_{|\Sigma} = \tau \kappa,
\end{equation*}
where $\kappa$ is the mean curvature of the surface $\Sigma$ and $\tau$ is the surface tension. The Hamiltonian is taken to be
\begin{equation*}
    H(v,\Sigma) = \frac{1}{2}\int_{D_{\Sigma}}\norm{v}^{2}\, dV + \tau \int_{\Sigma}dS.
\end{equation*}

However this bracket does not define a \emph{valid Hamiltonian structure} since it is not \emph{closed}. To show that, we will compute the bracket of two specific admissible functionals and show that the bracket is not an admissible functional. Let
\begin{equation*}
    F(v,\Sigma) = \frac{1}{2}\int_{D_{\Sigma}} f(\norm{v}^{2})\, dV, \qquad G(v,\Sigma) = \frac{1}{2}\int_{D_{\Sigma}} g(\norm{v}^{2})\, dV,
\end{equation*}
where $f$ and $g$ are smooth real functions. Those functionals are admissible and we have
\begin{equation*}
    \pdl{F}{v} = X_{f}, \qquad \pdl{F}{\Sigma} = \frac{1}{2}f(\norm{v}^{2})_{|\Sigma}, \qquad \pdl{G}{v} = X_{g}, \qquad \pdl{G}{\Sigma} = \frac{1}{2}g(\norm{v}^{2})_{|\Sigma},
\end{equation*}
where $X_{f}$ (resp. $X_{g}$) is the ($L^{2}$)-orthogonal projection of the vector field $f^{\prime}(\norm{v}^{2})v$ (resp. $g^{\prime}(\norm{v}^{2})v$ onto the space of divergence free vector fields.

\begin{proposition}
$\poisson{F}{G}$ is not an admissible function.
\end{proposition}

\begin{proof}
We have
\begin{align*}
    H(v,\Sigma) & = \poisson{F}{G}(v,\Sigma) \\
      & = \int_{D_{\Sigma}} \curl v \cdot (X_{f}\times X_{g})\, dV + \frac{1}{2} \int_{\Sigma} \left\{f(\norm{v}^{2})(X_{g}\cdot n) - g(\norm{v}^{2})(X_{f}\cdot n)\right\}dS \\
      & = \int_{D_{\Sigma}} \curl v \cdot (X_{f}\times X_{g})\, dV + \int_{D_{\Sigma}} \left\{f^{\prime}(\norm{v}^{2})(v \cdot X_{g}) - g^{\prime}(\norm{v}^{2})(v \cdot X_{f})\right\}\, dV.
\end{align*}
Let's denote the first integral in this expression by $H_{1}$ and the second one by $H_{2}$. We have
\begin{multline*}
    D_{v}H_{2}\cdot \delta v = \frac{1}{2} \int_{D_{\Sigma}} \Big\{ \left(\pdd{f}{v}\cdot \delta v\right)\cdot X_{g} -  \left(\pdd{g}{v}\cdot \delta v\right)\cdot X_{f} \\
    + \pd{f}{v} \cdot (D_{v}X_{g}\cdot \delta v) - \pd{g}{v} \cdot (D_{v}X_{f}\cdot \delta v) \Big\} dV
\end{multline*}
which can be rewritten as
\begin{multline*}
    D_{v}H_{2}\cdot \delta v = \frac{1}{2} \int_{D_{\Sigma}} \Big\{ \left(\pdd{f}{v}\cdot X_{g}\right)\cdot \delta v -  \left(\pdd{g}{v}\cdot X_{f}\right)\cdot \delta v \\
    + (D_{v}X_{g}\cdot X_{f})\cdot \delta v - (D_{v}X_{f}\cdot X_{g}) \cdot \delta v \Big\} dV,
\end{multline*}
using the property of symmetry of second Fr\'{e}chet derivative. That is the partial Fr\'{e}chet derivative of $H_{2}$ relative to $v$ admit a gradient. Therefore, this will be the case for $H$ if and only if this is true for $H_{1}$. We have
\begin{multline*}
    D_{v}H_{1}\cdot \delta v = \int_{D_{\Sigma}} \curl (\delta v) \cdot (X_{f}\times X_{g})\, dV \\
    + \int_{D_{\Sigma}} \curl v \cdot ([D_{v}X_{f}\cdot \delta v]\times X_{g})\, dV + \int_{D_{\Sigma}} \curl v \cdot (X_{f}\times [D_{v}X_{g}\cdot \delta v])\, dV.
\end{multline*}
In this expression, the last two terms are of gradient type because of the symmetry of the second Fr\'{e}chet derivative. The first term can be rewritten as
\begin{equation*}
    \int_{D_{\Sigma}} \left\{\delta v \cdot \curl (X_{f}\times X_{g}) + \dive(\delta v \times  (X_{f}\times X_{g}))\right\}\, dV,
\end{equation*}
which is definitely not of gradient type because of the divergence term. This achieves the proof that $\poisson{F}{G}$ is not an admissible function.
\end{proof}

\begin{remark}
The second LMMR bracket presented in Section~\ref{subsec:second_LMMR} can also be defined for free boundary problems with the same difficulties, that is the necessity of a non trivial boundary condition in the definition of admissible functionals. Is it possible to define a \emph{usable} and \emph{valid} bracket for free boundary problems using the method of Soloviev~?
\end{remark}

% ----------------------------------------------------------------


\begin{thebibliography}{10}

\bibitem{Arn65}
V.~I. Arnold, \emph{On conditions for non-linear stability of plane stationary
  curvilinear flows of an ideal fluid}, Dokl. Akad. Nauk SSSR \textbf{162}
  (1965), 975--978. \MR{MR0180051 (31 \#4288)}

\bibitem{Arn66}
\bysame, \emph{Sur la g{\'e}om{\'e}trie diff{\'e}rentielle des groupes de {L}ie
  de dimension infinie et ses applications {\`a} l'hydrodynamique des fluides
  parfaits}, Ann. Inst. Fourier (Grenoble) \textbf{16} (1966), no.~1, 319--361.
  \MR{34 \#1956}

\bibitem{BFLS98}
G.~Barnich, R.~Fulp, T.~Lada, and J.~Stasheff, \emph{The sh {L}ie structure of
  {P}oisson brackets in field theory}, Comm. Math. Phys. \textbf{191} (1998),
  no.~3, 585--601. \MR{MR1608547 (99j:17030)}

\bibitem{CH93}
R.~Camassa and D.D. Holm, \emph{An integrable shallow water equation with
  peaked solitons}, Phys. Rev. Lett. \textbf{71} (1993), no.~11, 1661--1664.
  \MR{MR1234453 (94f:35121)}

\bibitem{Con06}
A.~Constantin, \emph{The trajectories of particles in {S}tokes waves}, Invent.
  Math. \textbf{166} (2006), no.~3, 523--535. \MR{MR2257390}

\bibitem{CE07}
A.~Constantin and J.~Escher, \emph{Particle trajectories in solitary water
  waves}, Bull. Amer. Math. Soc. \textbf{44} (2007), 423--431.

\bibitem{CK03}
A.~Constantin and B.~Kolev, \emph{Geodesic flow on the diffeomorphism group of
  the circle}, Comment. Math. Helv. \textbf{78} (2003), no.~4, 787--804.

\bibitem{CK06}
\bysame, \emph{Integrability of invariant metrics on the diffeomorphism group
  of the circle}, J. Nonlinear Sci. \textbf{16} (2006), no.~2, 109--122.

\bibitem{CSW06}
A.~Constantin, D.~H. Sattinger, and W.~Strauss, \emph{Variational formulations
  for steady water waves with voticity}, J. Fluid Mech. \textbf{548} (2006),
  151--163. \MR{MR2264220}

\bibitem{CS04}
A.~Constantin and W.~Strauss, \emph{Exact steady periodic water waves with
  vorticity}, Comm. Pure Appl. Math. \textbf{57} (2004), no.~4, 481--527.
  \MR{MR2027299 (2004i:76018)}

\bibitem{Gar71}
C.~S. Gardner, \emph{Korteweg-de {V}ries equation and generalizations. {IV}.
  {T}he {K}orteweg-de {V}ries equation as a {H}amiltonian system}, J.
  Mathematical Phys. \textbf{12} (1971), 1548--1551. \MR{MR0286402 (44 \#3615)}

\bibitem{GD79}
I.~M. Gel'fand and I.~J. Dorfman, \emph{Hamiltonian operators and algebraic
  structures associated with them}, Funktsional. Anal. i Prilozhen. \textbf{13}
  (1979), no.~4, 13--30, 96. \MR{MR554407 (81c:58035)}

\bibitem{GD81}
\bysame, \emph{Hamiltonian operators and infinite-dimensional {L}ie algebras},
  Funktsional. Anal. i Prilozhen. \textbf{15} (1981), no.~3, 23--40.
  \MR{MR630337 (82j:58045)}

\bibitem{Kol04}
B.~Kolev, \emph{Lie groups and mechanics: an introduction}, J. Nonlinear Math.
  Phys. \textbf{11} (2004), 480--498.

\bibitem{Kol07}
\bysame, \emph{Bi-hamiltonian systems on the dual of the lie algebra of vector
  fields of the circle and periodic shallow water equations}, A para{\^\i}tre
  dans Phil. Trans. R. Soc. A, 2006.

\bibitem{KS06}
B.~Kolev and D.~H. Sattinger, \emph{Variational principles for water waves},
  SIAM J. Math. Anal. \textbf{38} (2006), no.~3, 906--920 (electronic).
  \MR{MR2262948}

\bibitem{LMMR85}
D.~Lewis, J.~Marsden, R.~Montgomery, and T.~Ratiu, \emph{The {H}amiltonian
  structure for dynamic free boundary problems}, Phys. D \textbf{18} (1986),
  no.~1-3, 391--404, Solitons and coherent structures (Santa Barbara, Calif.,
  1985). \MR{MR838352 (87m:22053)}

\bibitem{MW74}
J.~Marsden and A.~Weinstein, \emph{Reduction of symplectic manifolds with
  symmetry}, Rep. Mathematical Phys. \textbf{5} (1974), no.~1, 121--130.
  \MR{MR0402819 (53 \#6633)}

\bibitem{Mis98}
G.~Misio{\l}ek, \emph{A shallow water equation as a geodesic flow on the
  {B}ott-{V}irasoro group}, J. Geom. Phys. \textbf{24} (1998), no.~3, 203--208.
  \MR{MR1491553 (99d:58018)}

\bibitem{Olv93}
P.~J. Olver, \emph{Applications of {L}ie groups to differential equations},
  second ed., Graduate Texts in Mathematics, vol. 107, Springer-Verlag, New
  York, 1993. \MR{MR1240056 (94g:58260)}

\bibitem{Sol93}
V.~O. Soloviev, \emph{Boundary values as {H}amiltonian variables. {I}. {N}ew
  {P}oisson brackets}, J. Math. Phys. \textbf{34} (1993), no.~12, 5747--5769.
  \MR{MR1246246 (94i:58072)}

\bibitem{Sol97b}
\bysame, \emph{Difference between admissible and ``differentiable''
  {H}amiltonians}, Phys. Rev. D (3) \textbf{55} (1997), no.~12, 7973--7976.
  \MR{MR1455124 (98e:58082)}

\bibitem{Sol97a}
\bysame, \emph{Divergences in formal variational calculus and boundary terms in
  {H}amiltonian formalism}, Symplectic singularities and geometry of gauge
  fields (Warsaw, 1995), Banach Center Publ., vol.~39, Polish Acad. Sci.,
  Warsaw, 1997, pp.~373--388. \MR{MR1458672 (98k:58076)}

\bibitem{Sol02a}
\bysame, \emph{Boundary values as {H}amiltonian variables. {II}. {G}raded
  structures}, J. Math. Phys. \textbf{43} (2002), no.~7, 3636--3654.
  \MR{MR1908690 (2003d:37123)}

\bibitem{Sol02b}
\bysame, \emph{Boundary values as {H}amiltonian variables. {III}. {I}deal fluid
  with a free surface}, J. Math. Phys. \textbf{43} (2002), no.~7, 3655--3675.
  \MR{MR1908691 (2003e:37091)}

\bibitem{Vai94}
I.~Vaisman, \emph{Lectures on the geometry of {P}oisson manifolds},
  Birkh{\"a}user Verlag, Basel, 1994. \MR{95h:58057}

\bibitem{Zak68}
V.~E. Zakharov, \emph{Stability of periodic waves of finite amplitude on the
  surface of a deep fluid}, J. Appl. Mech. Tech. Phys. \textbf{2} (1968),
  190--194.

\end{thebibliography}
\end{document}